\documentclass{iopart}

\usepackage{amsfonts}
\usepackage{setstack}
\usepackage{color}
\usepackage{cite}

\newcommand{\Snc}{S_{\mathrm{n.c.}}}
\newcommand{\dwedge}{\dot{\wedge}}

\newcommand{\JC}{J^{\mathrm{n.h.}}}

\newcommand{\oxi}{{\bar{\xi}}}
\newcommand{\txi}{{\tilde{\xi}}}

\newcommand{\Cartan}{\mathrm{Cartan}}

\usepackage{mathrsfs} 
\usepackage{latexsym} 
\usepackage{amssymb}

\usepackage{amsthm}

\newcommand{\calC}{\mathcal{C}}

\newcommand{\calE}{\mathcal{E}}

\newcommand{\calS}{\mathcal{S}}

\newcommand{\bfR}{\mathbb{R}}
\newcommand{\bfS}{\mathbb{S}}

\newcommand{\lie}{\mathscr{L}}

\newcommand{\pr}{\mathrm{pr}}

\newcommand{\dual}[1]{\left<{#1}\right>}

\newcommand{\im}{\mathrm{Im}\,}

\newcommand{\dd}[1]{\frac{\d}{\d #1}}

\newcommand{\dde}{\dd{\epsilon}}
\newcommand{\pdd}[1]{\frac{\partial}{\partial #1}}

\newcommand{\contraction}{\vrule height 0pt depth 0.4pt width 3pt
  \vrule height 7pt depth 0.4pt \kern 3pt}

\newcommand{\be}{\begin{equation}}
\newcommand{\ee}{\end{equation}}

\newtheoremstyle{jvk-thm} %
  {}{}{\itshape}{}{\bfseries}{.}{ }{}
\newtheoremstyle{jvk-rem} %
  {}{}{\upshape}{}{\bfseries}{.}{ }{}

\theoremstyle{jvk-thm}
\newtheorem{definition}{Definition}[section]
\newtheorem{lemma}[definition]{Lemma}
\newtheorem{theorem}[definition]{Theorem}

\newtheorem{corollary}[definition]{Corollary}

\newenvironment{mproof}{\textbf{Proof:}\,}{\hfill$\Box$}

\theoremstyle{jvk-rem}

\newtheorem{remarkth}[definition]{Remark}
\newenvironment{remark}{\begin{remarkth}}{\hfill$\diamond$\end{remarkth}}

\newenvironment{AlphaList}{%
  \begin{enumerate}}{\end{enumerate}}

\newcommand{\ol}[1]{\overline{#1}}
\newcommand{\ul}[1]{\underline{#1}}

\newcommand{\arxiv}[1]{\texttt{#1}}

\newcommand{\gothg}{\mathfrak{g}}

\newcommand{\gothX}{\mathfrak{X}}

\renewcommand{\d}{\mathrm{d}}

\newcommand{\qforall}{\quad \mbox{for all}\:}

\newcommand{\where}{\quad \mathrm{where}\:}
\newcommand{\qqand}{\quad \mathrm{and} \quad}

\begin{document}


\title{Symmetry aspects of nonholonomic field theories}
\author{Joris Vankerschaver$^{1,2}$, David Mart{\'\i}n de Diego$^3$}

\address{$^1$ Control and Dynamical Systems, California Institute of
  Technology, MC 107-81, Pasadena CA 91125}

\address{$^2$ Department of Mathematical Physics and Astronomy, Ghent
  University, Krijgslaan 281, B-9000 Ghent, Belgium}

\ead{jv@caltech.edu}

\address{$^3$ Instituto de Matem\'aticas y F{\'\i}sica Fundamental,
  Consejo Superior de Investigaciones Cient\'{\i}ficas, Serrano 123,
  28006 Madrid, Spain }

\ead{d.martin@imaff.cfmac.csic.es}

\begin{abstract}
  The developments in this paper are concerned with nonholonomic field
  theories in the presence of symmetries.  Having previously treated
  the case of vertical symmetries, we now deal with the case where the
  symmetry action can also have a horizontal component.  As a first
  step in this direction, we derive a new and convenient form of the
  field equations of a nonholonomic field theory.  Nonholonomic
  symmetries are then introduced as symmetry generators whose virtual
  work is zero along the constraint submanifold, and we show that for
  every such symmetry, there exists a so-called momentum equation,
  describing the evolution of the associated component of the momentum
  map.  Keeping up with the underlying geometric philosophy, a small
  modification of the derivation of the momentum lemma allows us to
  treat also generalized nonholonomic symmetries, which are vector
  fields along a projection.  Such symmetries arise for example in
  practical examples of nonholonomic field theories such as the
  Cosserat rod, for which we recover both energy conservation (a
  previously known result), as well as a modified conservation law
  associated with spatial translations.
\end{abstract}

\maketitle


\section{Introduction}

Knowledge of symmetry is fundamental to the understanding of
mechanical systems and field theories, whether classical or not.
Results such as Noether's theorem, the Marsden-Weinstein reduction
procedure, etc. all attest to this.  In the presence of nonholonomic
constraints, however, the concept of symmetry becomes somewhat more
subtle: Noether's theorem (see \cite{symm04} and the references
therein) for example no longer holds automatically.

Surprisingly, there is a more sophisticated analogue of Noether's
theorem, going by the name of the \emph{nonholonomic momentum
  equation} \cite{bloch96, reduction99}, which plays a fundamental
role in the reduction procedure for general nonholonomic systems.
This equation describes the evolution of the components of the
momentum map under the nonholonomic flow and its derivation relies
heavily on the use of differential geometry for the formulation of the
nonholonomic problem.

In the case of \emph{classical field theories}, a similar result was
derived in \cite{Vankerschaver05} for a restricted class of
symmetries, namely those symmetries whose infinitesimal vector field
is vertical, or alternatively, symmetries which act trivially on the
space of independent variables of the theory.  Nevertheless, many
physically interesting symmetries are not vertical, and an extension
of this result to a more general class of symmetries is therefore
needed: this is the subject of the present paper.  Examples of
non-vertical symmetries include translation in time, yielding
conservation of energy, without doubt the prime example of a conserved
quantity.

\subsection*{Outline of this paper}

In this paper, we derive a generalized form of the nonholonomic
momentum equation for two classes of symmetries.  The first consists
of symmetries which have a nontrivial horizontal component, while for
the second class we restrict our attention to vertical symmetries
whose infinitesimal generator is a generalized vector field,
\emph{i.e.} a vector field along the projection $\pi_{1, 0}$.  More
information on the application of generalized symmetries to
differential equations can be found in \cite{Olver}.

In section~\ref{sec:1stFE} we derive a new form of the field equations
for a classical field theory with nonholonomic constraints, which is
particularly well-suited to the derivation of the momentum equation in
section~\ref{sec:nhlemma}.  These equations are equivalent to the
nonholonomic field equations derived, for example, in
\cite{nhfields02, KrupkovaVolny, nhfields05} but they do not involve
the Lagrange multipliers, and are therefore closer in spirit to
H\"older's equations (see \cite{neimark}).  We derive the nonholonomic
momentum equation for two distinct classes of symmetries: firstly, for
the case of a symmetry group whose infinitesimal generators are
\emph{projectable} vector fields on the total space, and secondly, for
a symmetry group acting \emph{vertically} on the total space, but
whose associated nonholonomic symmetries are \emph{generalized vector
fields}. 

This theory is applied in section~\ref{sec:examples} to a number of
examples.  In section~\ref{sec:nonlin} we obtain a new class of
nonholonomic symmetries for mechanical systems with nonlinear
constraints; Benenti's nonholonomic system (see \cite{benenti}) is
chosen as a straightforward but relevant example and a number of new
conservation laws are derived.  In section~\ref{nonhcoss} we derive
nonholonomic conservation laws for the nonholonomic Cosserat rod, an
example of a nonholonomic field theory (see \cite{nonhcosserat}).
These conservation laws are associated to translations in time and
spatial translations, respectively.  The former is a horizontal
symmetry, while the latter is described by a (vertical) generalized
vector field.

Finally, in the appendix, we elaborate somewhat further on the choice
of reaction forces for a nonholonomic field theory.  As the material
in this section is not crucial to the developments in the remainder of
the paper, we have delegated it to the appendix.

\subsection*{Notation}

Throughout this paper, all geometric objects are assumed to be smooth.
We will denote the contraction of a vector field $V$ and a form
$\alpha$ both as $i_V \alpha$ or as $V \contraction \alpha$, and we
will use these two notations interchangeably.  The Lie derivative is
denoted by $\lie$.  We will frequently use the notation
$\bigwedge^k(M)$, where $M$ is a manifold, to denote the $k$-fold
exterior product of $TM$ with itself.  Thus, $\bigwedge^k(M)$ is a
bundle over $M$ whose sections are $k$-forms on $M$.

\section{Classical field theories}

As is customary in the geometric treatment of classical field theories
(see \cite{CrampinMS, gimmsyI, symm04, Saunders89} and the references
therein), fields are modelled as sections of a fibre bundle $\pi : Y
\rightarrow X$ of rank $m$, where the base space $X$ is an $(n +
1)$-dimensional oriented manifold (with volume form $\eta$), and the
total space $Y$ has dimension $n + m + 1$.  Local coordinates on $X$
are denoted by $(x^\mu)$, $\mu = 0, \ldots, n$ and are supposed to be
such that the volume form $\eta$ can be locally written as
\[
  \eta = \d^{n+1}x := \d x_0 \wedge \cdots \wedge \d x_n.
\]
In addition, we assume a local system of bundle coordinates $(x^\mu,
y^a)$, $\mu = 0, \ldots, n$, $a = 1, \ldots, m$, on $Y$ to be given.

Over $Y$, there exists a tower of jet bundles
\[
   \cdots \longrightarrow J^2\pi
   \stackrel{\pi_{2,1}}{\longrightarrow} J^1 \pi
   \stackrel{\pi_{1,0}}{\longrightarrow} Y.
\]
The elements of $J^k \pi$ are equivalence classes of local sections of
$\pi$, where two sections are said to be equivalent at a point $x$ of
$X$ if their $k$th-order Taylor expansions agree at that point.  We
denote the equivalence class of a section as $j^k_x \phi$.  More
information on jet bundles can be found in \cite{Saunders89}.  We will
mostly only need the first-order jet bundle $J^1\pi$; the second-order
jet bundle will make a brief appearance in section~\ref{nonhcoss} (see
for instance \cite{LeRo} for a review of mechanics on higher order jet
bundles).

We will need as basic geometric tools the concepts of contact forms
and jet prolongation of vector fields. A contact $m$-form on $J^k \pi$
will be any $m$-form $\theta$ satisfying $(j^k\phi)^*\theta$ for every
local section $\phi$ of $\pi$. The set of contact 1-forms on $J^1\pi$
defines a distribution ${\mathcal D}_{\Cartan}$, called the Cartan
distribution.  Using this distribution, it is easy to define the 1-jet
prolongation of a vector field $\xi_Y$ on $Y$ as the unique vector
field $j^1 \xi_Y$ projectable on $\xi_Y$ and preserving the Cartan
distribution, \emph{i.e.} $\lie_{j^1 \xi_{Y}}{\mathcal
  D}_{\Cartan}\subseteq{\mathcal D}_{\Cartan}$.  In coordinates, if 
  $\xi_Y = \xi^{\mu}(x, y) \pdd{x^\mu} + \xi^a(x, y) \pdd{y^a}$, 
then 
\[\fl
  j^1 \xi_Y = \xi^{\mu} \pdd{x^\mu} + \xi^a \pdd{y^a}
    + \left( \frac{\d \xi^a}{\d x^\nu} - y^a_\mu \frac{\partial
      \xi^\mu}{\partial x^\nu} \right) \pdd{y^a_\nu}, 
\quad \text{where } 
 \frac{\d}{\d x^\nu} = \frac{\partial}{\partial x^\nu} +
y^a_\nu \pdd{y^a}.
\]

If we adopt the coordinate systems on $X$ and $Y$ as above, then
$J^1\pi$ is equipped with an induced coordinate system $(x^\mu, y^a;
y^a_\mu)$, $\mu = 0, \ldots, n$, $a = 1, \ldots, m$.  With respect to
this coordinate system, the projection $\pi_{1, 0}: J^1\pi \rightarrow
Y$ is given by $\pi_{1, 0}(x^\mu, y^a; y^a_\mu) = (x^\mu, y^a)$.

For later use, we mention the existence of a distinguished
vector-valued $(n+1)$-form $S_\eta$ on $J^1\pi$, which is called the
\emph{vertical endomorphism}.  In coordinates, $S_\eta$ reads
\begin{equation} \label{vertend}
  S_\eta = (\d y^a - y^a_\nu \d x^\nu) \wedge \d^{n}x_\mu
    \otimes \frac{\partial}{\partial y^a_\mu}\,,
\end{equation}
where
\[\
  \d^n x_\mu :=
    \frac{\partial}{\partial x^{\mu}} \contraction \d^{n + 1}x.
\]

For the purpose of this paper, a Lagrangian will be a function $L$ on
$J^1\pi$.  Given a Lagrangian $L$, we define the Poincar\'e-Cartan
$(n + 1)$-form $\Theta_L$ as
\[
\Theta_L := S^\ast_\eta (\d L) + L \eta = \frac{\partial L}{\partial
  y^a_\mu}(\d y^a - y^a_\nu \d x^\nu) \wedge \d^{n}x_\mu + L \d^{n +
  1}x.
\]
We also put $\Omega_L := -\d \Theta_L$, to which we refer as the
Poincar\'e-Cartan $(n+2)$-form.

For more information on the geometry of jet bundles, see
\cite{Saunders89}.

\subsection*{The Euler-Lagrange equations}


Let there be given a Lagrangian $L$.  The dynamics of the field theory
is described by the Euler-Lagrange equations associated to $L$; they
express that the field is an extremum of the following action
functional:
\begin{equation} \label{action}
S(\phi) = \int_X L\left(x^\mu, \phi^a(x), \frac{\partial
    \phi^a}{\partial x^\mu}\right) \d^{n+1}x.
\end{equation}

By varying the action $S$ with respect to a variation $V$, we obtain
after integrating by parts
\begin{equation} \label{varieren}
  0 = \dde S(j^1 (\Phi_\epsilon \circ \phi) )\Big|_{\epsilon = 0} =
    \int_U \left( \frac{\partial L}{\partial y^a} - \frac{\d}{\d
        x^\mu} \frac{\partial L}{\partial y^a_\mu} \right) V^a
    \d^{n+1}x,
\end{equation}
where $\Phi_\epsilon$ is a finite variation associated to $V$.
Here, a \emph{variation} of a field $\phi$ over an open set $U
\subset X$ is a $\pi$-vertical vector field $V$ defined on an open
neighborhood of $\phi(U)$.  The associated finite variation
$\Phi_\epsilon$ is nothing but the flow of $V$.  Note that the
composition $\Phi_\epsilon \circ \phi$ is again a local section of
$\pi$.

In coordinates, a section $\phi$ of $\pi$ is an extremum of
(\ref{action}) if and only if it satisfies the familiar
Euler-Lagrange equations, given by
\[
  \pdd{x^\mu} \left( \frac{\partial L}{\partial
      y^a_\mu}(j^1\phi)\right) -
  \frac{\partial L}{\partial y^a}(j^1\phi) = 0.
\]
There exist various intrinsic formulations of these equations (see
\cite{overview02} for an overview), of which we mention just one.  It
can be shown by a straightforward coordinate calculation that the
Euler-Lagrange equations are equivalent to the following set of
intrinsic equations:
\[
  (j^1\phi)^\ast( i_W \Omega_L) = 0 \qforall W \in \gothX(J^1\pi).
\]

In what follows, we will be mostly interested in similar equations for
field theories which are subject to nonholonomic constraints.  This is
the subject of the next section.

\section{The nonholonomic field equations} \label{sec:1stFE}

A nonholonomic field theory is given by the specification of three
objects (see also \cite{nhfields05}):
\begin{enumerate}
  \item a Lagrangian $L: J^1\pi \rightarrow \bfR$;
  \item a constraint submanifold $\calC \hookrightarrow J^1\pi$, such
    that the restriction of the projection $(\pi_{1, 0})_{|\calC}$
    defines a subbundle of $\pi_{1, 0}: J^1\pi \rightarrow Y$;
  \item a bundle of reaction forces $F$, where the elements $\Phi$ of
    $F$ are $(n + 1)$-forms defined along $\calC$, \emph{i.e.} maps
    from $\calC \subset J^1\pi$ to $\Omega^{n+1}(J^1\pi)$.  The
    elements of $F$ have to satisfy the following requirements:
    \begin{enumerate}
    \item $\Phi$ is $n$-horizontal, \emph{i.e.} $\Phi$ vanishes when
      contracted with any two $\pi_1$-vertical vector fields;
    \item $\Phi$ is $1$-contact, \emph{i.e.} $(j^1\phi)^*\Phi = 0$ for
      any section $\phi$ of $\pi$.
    \end{enumerate}
    It can be shown that any element $\Phi$ of $F$ is of the following
    form:
    \begin{equation} \label{locform}
      \Phi = A^{\mu}_a (\d y^a - y^a_\nu \d x^\nu) \wedge \d^nx_\mu,
    \end{equation}
    where $A^{\mu}_a$ are functions on $J^1\pi$.
\end{enumerate}

For the sake of simplicity, we will assume $\calC$ to be defined by
the vanishing of $k$ functionally independent functions
$\varphi^\alpha$ on $J^1\pi$.  Furthermore, we will assume that $F$ is
globally generated by $l$ generators $\Phi^\kappa$ of the following
form:
\[
      \Phi^\kappa = A^{\kappa \mu}_a (\d y^a - y^a_\nu \d x^\nu)
      \wedge \d^nx_\mu \quad (\kappa = 1, \ldots, l).
\]
In practice, the dimension $l$ of $F$ will be equal to the codimension
$k$ of $\calC$.  There seems to be no a priori reason for supposing
that $k = l$.  In most cases, however, $F$ will be determined by
$\calC$ through application of the Chetaev principle, described in
remark~\ref{rem:chetaev} below.

In the nonholonomic treatment of constraints, a special role is played
by infinitesimal variations of the fields which are compatible with
the constraint, as in the following definition.

\begin{definition} \label{def:admvar} A variation $V$ of a field
  $\phi$ (taking values in $\calC$, \emph{i.e.} such that $j^1\phi \subset
  \calC$) defined over an open subset $U$ with compact closure is
  \emph{admissible} if
  \begin{equation} \label{admvar}
   (j^1\phi)^\ast (j^1 V \contraction \Phi) = 0 \qforall \Phi \in F.
  \end{equation}
\end{definition}


By varying the action $S$ with respect to an admissible variation $V$,
we obtain after integrating by parts again (\ref{varieren}).  If the
variations $V$ were arbitrary, then this would immediately yield the
Euler-Lagrange equations.  However, this is not the case as the
variations have to be admissible.  Additional \emph{reaction forces}
will therefore appear in the Euler-Lagrange equations, whose role it
is to constrain the solution $\phi$ to the constraint submanifold.

\begin{definition}
  A local section $\phi$ of $\pi$, defined on an open subset $U
  \subset X$ with compact closure, is a \emph{solution} of the
  nonholonomic problem determined by $L$, $\calC$, and $F$ if
  $j^1\phi(U) \subset \calC$ and (\ref{varieren}) holds for all
  admissible variations $V$ of $\phi$.
\end{definition}

It follows from (\ref{varieren}) that a local section $\phi$ is a
solution of the nonholonomic problem if it satisfies the
\emph{nonholonomic Euler-Lagrange equations}:
\begin{equation} \label{eq:nh} \left[ \frac{\partial L}{\partial y^a}
    - \frac{\d}{\d x^\mu} \frac{\partial L}{\partial y^a_\mu}
  \right](j^2\phi) = \lambda_{\alpha\kappa}
  A^{\alpha\kappa}_a(j^1\phi) \qqand \varphi^\alpha(j^1\phi) = 0.
\end{equation}
Here, $\lambda_{\alpha\kappa}$ are unknown Lagrange multipliers, to be
determined from the constraints.  An intrinsic form of these equations
is derived below in theorem~\ref{thm:EL}, but first we need the
following technical results.

\begin{lemma}[lemma~3.2 in \cite{gimmsyI}] \label{lemma:gimmsy}
  Let $W$ be a vector field on $J^1\pi$.  If $\phi$ is a section of
  $\pi$ and if either $W$ is tangent to the image of $j^1 \phi$ or if
  $W$ is $\pi_{1, 0}$-vertical, then $(j^1 \phi)^\ast(i_W \Omega_L) =
  0$.
\end{lemma}

Now, let $\phi$ be a section such that the image of $j^1 \phi$ is
a subset of $\calC$ and consider a vector field $W$ which is
tangent to the image of $j^1\phi$, \emph{i.e.} there exists a
vector field $w$ on $X$ such that $T_x j^1\phi (w(x)) = W( j^1_x
\phi)$ for all $x \in X$.  One can follow a similar reasoning as
in the proof of lemma~3.2 in \cite{gimmsyI} to show that
\[
  (j^1 \phi)^\ast ( W \contraction \Phi ) = w \contraction ((j^1
  \phi)^\ast \Phi )
\]
for any $\Phi \in F$.  Since $\Phi$ is $1$-contact, the right-hand
side of this expression vanishes.  On the other hand, if $W$ is
$\pi_{1, 0}$-vertical, it follows automatically that $(j^1\phi)^\ast (
W \contraction \Phi ) = 0$.  We have therefore proved the following
lemma:

\begin{lemma} \label{lemma:moi} Let $\phi$ be a section of $\pi$ such
  that $j^1_x\phi \in \calC$ for all $x \in U \subset X$.  If either
  $W$ is tangent to the image of $j^1\phi$ or $W$ is $\pi_{1,
    0}$-vertical, then $(j^1\phi)^\ast ( W \contraction \Phi ) = 0$
  for all $\Phi \in F$.
\end{lemma}

Henceforth, we shall call any vector field $W$ on $J^1\pi$
\emph{admissible} with respect to a section $\phi$ of $\pi$ if $(j^1
\phi)^\ast (W \contraction \Phi) = 0$ for all $\Phi \in F$.

\begin{theorem} \label{thm:EL}
  Let $\phi$ be a section of $\pi$. If $\im j^1\phi \subset \calC$, then
  the following assertions are equivalent:
  \begin{AlphaList}
    \item $\phi$ is a stationary point of the action (\ref{action})
      under admissible variations; \label{eq1}
    \item $\phi$ satisfies the Euler-Lagrange equations
      (\ref{eq:nh}); \label{eq2}
    \item for all vector fields $W$ on $J^1\pi$ such that
      $(j^1\phi)^\ast(W \contraction \Phi) = 0$ for all $\Phi \in F$,
      \label{eq3}
      \begin{equation} \label{eq:intrinsic}
        (j^1\phi)^\ast (W \contraction \Omega_L) = 0.
      \end{equation}
  \end{AlphaList}
\end{theorem}
\begin{mproof}
Let us first prove the equivalence of (\ref{eq1}) and (\ref{eq3}).
For arbitrary, not necessarily admissible variations, the following
result holds (this is equation~3C.5 in \cite{gimmsyI}):
\[
  \frac{\d}{\d\epsilon} S(\phi_\epsilon)\Big|_{\epsilon = 0}
    = -\int_U (j^1 \phi)^\ast(j^1 V \contraction \Omega_L).
\]
For admissible variations, from hypothesis (a), we have
\[
  \int_U (j^1 \phi)^\ast(j^1 V \contraction \Omega_L) = 0.
\]
Now, we may multiply $V$ by an arbitrary function on $X$ and this
result will still hold true.  The fundamental lemma of the calculus of
variations therefore shows that
\begin{equation} \label{eqIntrNH}
  (j^1 \phi)^\ast(j^1 V \contraction \Omega_L) = 0,
\end{equation}
for all admissible variations $V$ defined over $U$.  By using a
partition of unity as in \cite{gimmsyI}, it can then be shown that
(\ref{eqIntrNH}) holds for all $\pi$-vertical vector fields $V$ such
that $(j^1 V) \contraction \Phi = 0$ for all $\Phi \in F$.  This
expression is equivalent to (\ref{eq:intrinsic}): to see this, take an
arbitrary vector field $W$ on $J^1\pi$ such that $(j^1 \phi)^\ast(W
\contraction \Phi) = 0$ for all $\Phi \in F$.  The vector field $W$
can be decomposed as the following sum (to be considered along the
image of $j^1\phi$):
\[
  W = w_\Vert + j^1 V + v_{\pi_{1,0}},
\]
where $w_\Vert$ is tangent to the image of $j^1\phi$, $j^1V$ is the
prolongation of a $\pi$-vertical vector field $V$, and $v_{\pi_{1,
    0}}$ is a $\pi_{1, 0}$-vertical vector field.  Using
lemma~\ref{lemma:moi}, we have that
\[
(j^1\phi)^\ast( j^1V \contraction \Phi) = (j^1\phi)^\ast (W \contraction
\Phi) = 0,
\]
and from lemma~\ref{lemma:gimmsy}, we get $(j^1\phi)^\ast(W
\contraction \Omega_L) = (j^1\phi)^\ast(j^1 V \contraction \Omega_L)$.
The right-hand side of this equation vanishes since $j^1V$ is
admissible, and therefore we conclude that $W$ satisfies
(\ref{eq:intrinsic}).

The equivalence of (\ref{eq2}) and (\ref{eq3}) is just a matter of
writing out the definitions.  In coordinates, the left-hand side of
(\ref{eq:intrinsic}) reads (for a prolongation of a vertical vector
field $V$)
\[
(j^1 \phi)^\ast(j^1V \contraction \Omega_L) = V^a\left( \frac{\partial
    L}{\partial y^a}(j^1\phi) - \frac{\partial}{\partial x^\mu}
  \frac{\partial L}{\partial y^a_\mu}(j^1\phi) \right) \d^{n+1}x,
\]
and this holds for all admissible variations $V$.  Therefore, if
$\phi$ satisfies (\ref{eq:intrinsic}), then there exist functions
$\lambda_{\alpha\kappa}$ such that
\[
\left[\frac{\partial L}{\partial y^a} - \frac{\d}{\d x^\mu}
  \frac{\partial L}{\partial y^a_\mu}\right](j^2\phi) =
  \lambda_{\alpha\kappa}
A^{\alpha\kappa}_a(j^1\phi).
\]
The converse is similar.
\end{mproof}

We see from the proof of this theorem that only vertical vector fields
yield nontrivial results for (\ref{eq:intrinsic}).

\begin{remark} \label{rem:chetaev} The bundle of reaction forces $F$
  is commonly derived from the constraint submanifold $\calC$ through
  application of the Chetaev principle (see \cite{neimark}, as well as
  \cite{nhfields05} for an extension to the case of field theories).
  If the constraint submanifold is given as the zero level set of
  functions $\varphi^\alpha$, then according to this principle, $F$ is
  locally generated by the following forms:
  \[
  \Phi^\alpha := S^\ast_\eta (\d \varphi^\alpha) = \frac{\partial
    \varphi^\alpha}{\partial y^a_\mu} (\d y^a - y^a_\nu \d x^\nu)
  \wedge \d^n x_\mu.
  \]
  In the past, there has appeared some criticism over the use of the
  Chetaev principle (see \cite{marle98}), and as we shall see in the
  appendix, for classical field theories the Chetaev principle
  sometimes has to be modified.
\end{remark}

\section{The nonholonomic momentum equation} \label{sec:nhlemma}

In this section, we derive the nonholonomic momentum equation, the
nonholonomic counterpart to the well-known theorem of Noether.  More
in detail, we prove that for every \emph{nonholonomic symmetry} there
exists a certain partial differential equation, which reduces to a
conservation law when the constraints are absent.  In proving the
momentum lemma, many different starting assumptions can be made, and
we study two different setups:
\begin{enumerate}
\item in section~\ref{sec:horsymm}, we assume that the symmetry group
  $G$ acts by bundle automorphisms on $Y$.  In particular, we allow
  for the fact that $G$ acts nontrivially on $X$ as well.  In this
  case, nonholonomic symmetries are \emph{projectable} vector fields on
  $Y$.

\item in section~\ref{sec:versymm}, we model nonholonomic symmetries
  as vector fields along the projection $\pi_{1,0}$.  For the sake of
  simplicity, we assume in this case that the action of $G$ on $Y$ is
  \emph{vertical}, \emph{i.e.} $G$ acts trivially on $X$.  This setup
  is therefore complementary to the one described before, but this
  case can probably be extended even further.
\end{enumerate}

\subsection{The nonholonomic momentum map}

Let $G$ be a Lie group acting on $Y$ by bundle automorphisms;
\emph{i.e.} there exist smooth actions $\ol{\Phi}: G \times Y
\rightarrow Y$ and $\ul{\Phi}: G \times X \rightarrow X$ such that
$\pi ( \ol{\Phi}(g, y) ) = \ul{\Phi}(g, \pi(y))$ for all $g\in G$ and
$y \in Y$.  We use the following abbreviations: $\ol{\Phi}_g :=
\ol{\Phi}(g, \cdot)$ and $\ul{\Phi}_g := \ul{\Phi}(g, \cdot)$.
The Lie group $G$ acts on $J^1\pi$ by prolonged bundle automorphisms,
\emph{i.e.}
\[
j^1\Phi_g ( j^1_x \phi ) = j^1_x( \ol{\Phi}_g \circ \phi \circ
\ul{\Phi}_g^{-1}).
\]

Let us now assume that $G$ leaves invariant $L$, $\calC$, and $F$:
\[
  j^1\Phi_g^\ast(L\eta) = L\eta,  \quad
    j^1\Phi_g(\calC) \subset \calC \qqand
    (j^1\Phi_g)^\ast F \subset F
\]
for all $g \in G$.

We consider first the vector bundle $\gothg^F$ over $Y$, defined as
follows.  Denote by $\gothg^F(y)$ the linear subspace of $\gothg$
consisting of all $\xi \in \gothg$ such that
\begin{equation} \label{defbundle}
  j^1\xi_Y(\gamma) \contraction F = 0 \qforall \gamma \in \calC \cap
    \pi^{-1}_{1, 0}(y),
\end{equation}
where $\xi_Y$ is the infinitesimal generator of the action
corresponding to $\xi$, that is,
\[
\xi_Y(y)=\frac{\d}{\d t} \ol{\Phi}_{\exp(t\xi)} (y) \Big|_{t=0}.
\]
We assume that the disjoint union of all $\gothg^F(y)$, for all
$y \in Y$ can be given the structure of a vector bundle $\gothg^F$
over $Y$.  

To any section $\bar{\xi}$ of $\gothg^F$, one can associate a vector
field $\tilde{\xi}_Y$ on $Y$ according to the following prescription:
\begin{equation} \label{assocvf}
  \tilde{\xi}_Y(y) = \left[\bar{\xi}(y)\right]_Y(y).
\end{equation}

\begin{definition} \label{def:nhsymm}
  A \emph{nonholonomic symmetry} is a section $\bar{\xi}$ of
  $\gothg^F$ such that the associated vector field $\tilde{\xi}_Y$
  is $\pi$-projectable; \emph{i.e.} there exists a vector field
  $\tilde{\xi}_X$ on $X$ such that $T \pi \circ \tilde{\xi}_Y =
  \tilde{\xi}_X \circ \pi$.
\end{definition}

\begin{definition}\label{def:hsym}
A \emph{horizontal nonholonomic symmetry} is a section $\bar{\xi}$ of
$\gothg^F$ which is constant, \emph{i.e.} $\bar{\xi}(y) =
\bar{\xi}(y')$ for all $y, y' \in Y$.
\end{definition}

We may identify horizontal nonholonomic symmetries with elements
$\xi\in \gothg$ such that $\xi = \gothg^F(y)$ for all $y\in
Y$, that is,
\[  j^1\xi_Y(\gamma) \contraction F = 0 \qforall \gamma \in \calC \cap
    \pi^{-1}_{1, 0}(y) \quad \text{and for all} \quad y \in Y,
\]

We now define the \emph{nonholonomic momentum map} as the map
$\JC: \calC \rightarrow \bigwedge^n (J^1\pi) \otimes
\gothg^F$, constructed as follows.  Let $\bar{\xi}$ be any
section of $\gothg^F$ (for the construction of the momentum
map it does not matter whether the associated vector field is
projectable or not) and put
\begin{equation} \label{JC}
    \JC_\oxi = i_{j^1\txi_Y} \Theta_L,
\end{equation}
where $\txi$ is the vector field associated to $\bar{\xi}$ according
to (\ref{assocvf}).  Given this definition of $\JC_\oxi$, we define
$\JC$ by the following rule: 
\[
  \left< \JC, \bar{\xi} \right> = \JC_\oxi.
\]

If $\bar{\xi}$ is a nonholonomic symmetry, then the prolongation $j^1
\txi_Y$ of the associated vector field is admissible.  This is proved
below in corollary~\ref{corr:adm}.

\begin{lemma}
  Let $\bar{\xi}$ be a section of $\gothg^F$.  For a fixed  $y \in Y$, put
  $\xi := \bar{\xi}(y)$ and consider any $\gamma \in \pi_{1,
    0}^{-1}(y) \cap \calC$.  Then there exists a $\pi_{1, 0}$-vertical
  vector $v_\gamma \in T_\gamma J^1\pi$ such that
  \[
    j^1 \txi_Y (\gamma) = j^1 \xi_Y(\gamma) + v_\gamma.
  \]
\end{lemma}
\begin{mproof}
  Recall that $j^1 \xi_Y$ is the prolongation of the fundamental
  vector field $\xi_Y$ associated to the (fixed) Lie algebra element
  $\xi$.  The lemma follows from the fact that $T_\gamma \pi_{1,
    0}(j^1\txi_Y(\gamma)) = \txi_Y(y)$.  On the other hand,
  $T_{\gamma}\pi_{1,0}(j^1 \xi_Y(\gamma)) = \xi_Y(y)$, but by definition,
  $\txi_Y(y) = \xi_Y(y)$.
\end{mproof}

\begin{corollary} \label{corr:adm}
  Let $\bar{\xi}$ be a nonholonomic symmetry.  Then the prolonged
  vector field $j^1 \txi_Y$ is admissible with respect to any section
  $\phi$ of $\pi$.
\end{corollary}
\begin{mproof}
  Take $\gamma \in \calC$, put $y = \pi_{1, 0}(\gamma)$, and let $\xi
  = \bar{\xi}(y)$.  From definition~\ref{def:nhsymm}, we gather that
  for all $\Phi \in F$, $j^1 \xi_Y(\gamma) \contraction \Phi(\gamma) =
  0$.  But $j^1 \txi_Y(\gamma)$ is equal to $j^1 \xi_Y(\gamma)$ up to
  a vertical vector, while $\Phi \in F$ is semi-basic.  Therefore,
  \[
    j^1 \txi_Y(\gamma) \contraction \Phi(\gamma) =
  0.
  \]
  As this holds for all $\gamma \in \calC$, we conclude that
  $j^1\txi_Y$ is admissible with respect to any section.
\end{mproof}

\begin{remark}
  In the case of mechanical systems with linear constraints, the
  bundle $\gothg^F$, defined as above, coincides with the bundle
  $\calS$ used in \cite{bloch96}.
\end{remark}

\subsection{The nonholonomic momentum equation} \label{sec:horsymm}

\begin{theorem} \label{thm:nhmom}
  If $\phi$ is a solution of the nonholonomic field equations
  (\ref{eq:intrinsic}), then for any nonholonomic symmetry $\bar{\xi}$
  the associated component of the momentum map $\JC_\oxi$ satisfies
  the following \emph{nonholonomic momentum equation}:
  \[
  (j^1 \phi)^\ast (\d \JC_\oxi) = (j^1\phi)^\ast\left(\lie_{j^1
      \txi_Y}(L \eta)\right).
  \]
\end{theorem}
\begin{mproof}
  Let $\bar{\xi}$ be a nonholonomic symmetry.  Recall that the
  prolongation $j^1\txi$ of the associated vector field is admissible
  with respect to any section of $\pi$.  Therefore, since
  \[
\d \JC_\oxi = \lie_{j^1 \txi_Y} \Theta_L + i_{j^1 \txi_Y} \Omega_L,
  \]
  pulling back along a solution of the nonholonomic Euler-Lagrange
  equations gives us
  \begin{equation} \label{eq:derr} (j^1 \phi)^\ast (\d \JC_\oxi) =
    (j^1\phi)^\ast \left(\lie_{j^1 \txi_Y} \Theta_L\right).
  \end{equation}

  For the sake of notational convenience, let us say that two forms
  $\alpha$ and $\beta$ on $J^1\pi$ are equivalent (denoted by $\alpha
  \simeq \beta$) if they agree up to a contact form, \emph{i.e.}
  $\alpha \simeq \beta$ iff $\alpha = \beta + \theta$, where $\theta$
  is contact.  This is equivalent to saying that $(j^1\phi)^\ast
  \alpha = (j^1\phi)^\ast \beta$ for all sections $\phi$ of $\pi$.  We
  then have that
  \[
    \lie_{j^1 \txi_Y} \Theta_L =
      \lie_{j^1 \txi_Y}\left( S^\ast_\eta(\d L)  \right) +
      \lie_{j^1 \txi_Y}\left( L \eta \right).
  \]
  In coordinates, the first term on the right-hand side becomes
  \begin{eqnarray*}
    \lie_{j^1 \txi_Y}\left( S^\ast_\eta(\d L)  \right) & =
       \lie_{j^1 \txi_Y}\left( \frac{\partial L}{\partial y^a_\mu} (\d
         y^a - y^a_\nu \d x^\nu) \wedge \d^n x_\mu \right) \\
       & \simeq \frac{\partial L}{\partial y^a_\mu}
       \lie_{j^1 \txi_Y} (\d
         y^a - y^a_\nu \d x^\nu) \wedge \d^n x_\mu \\
       & = \frac{\partial L}{\partial y^a_\mu} \frac{\partial
         \xi^a}{\partial y^b} (\d y^b - y^b_\nu \d x^\nu) \wedge \d^n
       x_\mu \simeq 0
  \end{eqnarray*}
  and hence vanishes when pulled back along a prolongation of a
  section.  Here, we've used the fact that the prolongation $j^1
  \txi_Y$ can locally be written as follows:
  \[
  j^1 \txi_Y = \xi^\mu(x) \pdd{x^\mu} + \xi^a(x, y) \pdd{y^a}
    + \left( \frac{\d \xi^a}{\d x^\mu} - y^a_\nu \frac{\d \xi^\nu}{\d
        x^\mu} \right) \pdd{y^a_\mu},
  \]
  and that the Lie derivative of a contact form with respect to such a
  vector field is again a contact form.  Note especially that the
  coefficient $\xi^\mu(x)$ does not depend on $y$: this is a
  consequence of the projectability condition in
  definition~\ref{def:nhsymm}.
\end{mproof}
\begin{corollary}
If $\phi$ is a solution of the nonholonomic field equations
  (\ref{eq:intrinsic}), then for any horizontal nonholonomic symmetry $\xi$
  the   momentum map $\JC_\xi$ is a conservation law for the nonholonomic dynamics, that is,
      \[
  (j^1 \phi)^\ast (\d \JC_\xi) = 0.
  \]
\end{corollary}

\subsection{Vertical vector fields along the projection} \label{sec:versymm}

Not all nonholonomic symmetries can be modelled as sections of
$\gothg^F$; as we shall see, an important class consists of
sections of the pullback bundle $\pi^\ast_{1, 0} \gothg^F$,
\emph{i.e.} maps $\bar{\xi} : \calC \rightarrow \gothg^F$ such
that $\bar{\xi}(\gamma) \in \gothg^F(\pi_{1, 0}(\gamma))$ for
all $\gamma \in \calC$. Such a section can be represented as
\begin{equation} \label{nhsymm}
  \bar{\xi} = \xi^A(x^\mu, y^a, y^a_\mu) e_A,
\end{equation}
where $\xi^A(x^\mu, y^a, y^a_\mu)$ are locally defined functions on
$\calC$ and $\{e_A\}$ is a basis of $\gothg$.

For this treatment, we will make one important simplification:
\emph{we assume that the symmetry group $G$ acts vertically}.  It
should noted, though, that this is probably not a fundamental
restriction: it is likely that this approach can be generalized
further.

Our treatment is therefore somewhat restricted compared to the
previous paragraph.  On the other hand, by allowing as in
(\ref{nhsymm}) sections whose coefficients are functions on $\calC$
rather than just on $Y$, we are able to describe symmetries which do
not fit into the framework of the previous paragraph.  Examples will
be given in section~\ref{sec:examples}.

\begin{definition} \label{def:genns}
  A \emph{generalized nonholonomic symmetry} is a section $\bar{\xi}$
  of the pull-back bundle $\pi^\ast_{1, 0} \gothg^F$.
\end{definition}

A generalized nonholonomic symmetry $\bar{\xi}$ induces a vertical
vector field $\tilde{\xi}$ along $\pi_{1, 0}$, defined as follows: for
all $\gamma \in \calC$,
\[
  \txi(\gamma) =
\left[\bar{\xi}(\gamma)\right]_Y(y) \in T_y Y, \where y = \pi_{1,
  0}(\gamma).
\]
There exists a generalized notion of prolongation for vector fields
along $\pi_{1, 0}$ (see \cite{Saunders89}).  In coordinates, if $\txi
= \xi^a(x^\mu, y^b, y^b_\mu) \pdd{y^a}$, then
\[
  j^1\txi_Y = \xi^a \pdd{y^a} + \frac{\d \xi^a}{\d x^\mu}
  \pdd{y^a_\mu}, \where\
    \frac{\d \xi^a}{\d x^\mu} = \frac{\partial \xi^a}{\partial x^\mu}
    + \frac{\partial \xi^a}{\partial y^b} y^b_\mu + \frac{\partial
      \xi^a}{\partial y^b_\nu} y^b_{\mu\nu}.
\]
Note that $j^1\txi_Y$ is a vector field along $\pi_{2, 1}$, as its
coefficients depend on the second-order derivatives $y^b_{\mu\nu}$.

Having defined the prolongation $j^1\txi_Y$ as above, we now define the
associated component of the momentum map as
\[
  \JC_\oxi = i_{j^1\txi_Y} \Theta_L.
\]
This definition is formally identical to (\ref{JC}), but strictly
speaking, the momentum map is now an $n$-form along the projection
$\pi_{2, 1}$.  By pull-back, such a form induces an $n$-form on
$J^2\pi$.  Similarly, the concept of Lie derivation can be extended in
a natural fashion to the case of vector fields along the projection
(see \cite{Saunders89}).

\begin{theorem} \label{thm:nhmomgen}
  If $\phi$ is a solution of the nonholonomic field equations
  (\ref{eq:intrinsic}), then for any generalized nonholonomic symmetry
  $\bar{\xi}$ the associated component of the momentum map $\JC_\oxi$
  satisfies the following \emph{nonholonomic momentum equation}:
  \begin{equation} \label{momeq}
  (j^2 \phi)^\ast (\d \JC_\oxi) = (j^2\phi)^\ast\left(\lie_{j^1
      \txi_Y}(L \eta)\right).
  \end{equation}
\end{theorem}
\begin{mproof}
The proof is similar in spirit to that of theorem~\ref{thm:nhmom} but
there are some additional technical difficulties that need to be taken
into account.  For any solution $\phi$ of the nonholonomic field
equations, we have
\begin{equation} \label{eq:workout}
  (j^2 \phi)^\ast(\d \JC_\oxi) = (j^2 \phi)^\ast (\lie_{j^1 \txi_Y}
  \Theta_L) + (j^2 \phi)^\ast( i_{j^1 \txi_Y} \Omega_L ).
\end{equation}
For the first term on the right-hand side, one can check that
\[
   \lie_{j^1 \txi_Y} (\d y^a - y^a_\mu \d x^\mu) =
     \frac{\partial \xi^a}{\partial y^b}(\d y^b - y^b_\mu \d x^\mu)
     + \frac{\partial \xi^a}{\partial y^b_\mu} (\d y^b_\mu -
     y^b_{\mu\nu} \d x^\nu),
\]
a contact form on $J^2\pi$.  From this, it follows that the Lie
derivative of $\Theta_L$ is given by
\[
  (j^2\phi)^\ast(\lie_{j^1 \txi_Y} \Theta_L) =
  (j^2\phi)^\ast(\lie_{j^1 \txi_Y} (L \eta)).
\]

The second term on the right-hand side of (\ref{eq:workout}) is
zero. This is essentially a consequence of the fact that $\phi$ is a
solution of the field equations and can be proved using the following
observation.  Take $x \in X$ and consider a vector field $W$ on
$J^1\pi$ such that $W(j^1_x \phi) = j^1 \txi_Y(j^2_x \phi)$.  Then
$W(j^1_x \phi) \contraction \Phi = 0$ for all $\Phi \in F$, and
without loss of generality, we may choose $W$ to be such that this
equality holds in a neighbourhood of $j^1_x \phi$.  We now have that
\[
  (j^2 \phi)^\ast ( i_{j^1 \txi_Y} \Omega_L )(x) =
  (j^1 \phi)^\ast( i_W \Omega_L)(x).
\]
As $W$ is admissible with respect to $\phi$, we conclude that the
right-hand side is zero, and this in turn implies the vanishing of the
second term in (\ref{eq:workout}).
\end{mproof}




\section{Examples} \label{sec:examples}

\subsection{Mechanical systems with nonlinear
  constraints} \label{sec:nonlin}

As an illustrative example, we take a variation of Benenti's system
(see \cite{cortes, implicitnonhol} and the references therein), which
describes two point masses moving on a horizontal plane whose
velocities are constrained to be parallel.  The configuration space of
this system is $Q := \bfR^2 \times \bfR^2$, and so we put $X = \bfR$,
while $Y = \bfR \times Q$.  The projection $\pi$ is the projection
onto the first factor.  The Lagrangian for this system is
\[
  L = \frac{m}{2}(\dot{x}_1^2 + \dot{y}_1^2) +
  \frac{m}{2}(\dot{x}_2^2 + \dot{y}_2^2).
\]
Strictly speaking, $L$ is a function on $TQ$, but since $J^1 \pi$ is
isomorphic to $\bfR \times TQ$, we view $L$ as a (time-independent)
Lagrangian on $J^1\pi$.  The constraint is
\[
  \varphi \equiv \dot{x}_1 \dot{y}_2 - \dot{x}_2 \dot{y}_1 = 0.
\]
This is a nonlinear constraint, and determines a submanifold $\calC$
of $J^1\pi$.  If we assume Chetaev's principle to hold, the bundle of
reaction forces $F$ along $\calC$ is generated by the following
one-form:
\[
\Phi = \dot{y}_2 \theta_{x_1} + \dot{x}_1 \theta_{y_2} - \dot{y}_1
\theta_{x_2} - \dot{x}_2 \theta_{y_1},
\]
where the contact one-forms $\theta_{q^a}$ are defined as
$\theta_{q^a} = \d q^a - \dot{q}^a \d t$.

Consider now the obvious action of the Abelian group $\bfR^4$ on $Y$
by translations: $(a_1, b_1, a_2, b_2) \cdot (t; x_1, y_1, x_2, y_2) =
(t; x_1 + a_1, y_1 + b_1, x_2 + a_2, y_2 + b_2)$.  This action
fulfills all the necessary conditions needed for the momentum lemma:
it is vertical, and leaves invariant the Lagrangian $L$, the
constraint submanifold $\calC$, and the bundle of reaction forces $F$.

It is easy to check that the contraction of a vector field
$\tilde{\xi}$ of the following form
\[
  \tilde{\xi} = \alpha \dot{x}_1 \pdd{x_1} + \beta \dot{y}_1 \pdd{y_1}
    + \gamma \dot{x}_2 \pdd{x_2} + \delta \dot{y}_2 \pdd{y}_2 \quad
    (\alpha, \beta, \gamma, \delta \in \bfR)
\]
with $\Phi$ will vanish along $\calC$ if $\alpha + \delta = \beta +
\gamma$.  In this case, $\tilde{\xi}$ is a generalized nonholonomic
symmetry of the kind defined in definition~\ref{def:genns}.  (Strictly
speaking, the term ``generalized nonholonomic symmetry'' refers to the
section $\bar{\xi} := (\alpha \dot{x}_1, \beta \dot{y}_1, \gamma
\dot{x}_2, \delta \dot{y}_2)$ of $\pi^\ast_{1, 0} \gothg^F$.)

For the sake of convenience, let us take $\alpha = \beta = 1$ and
$\gamma = \delta = 0$.  The other cases are similar and the results
are summarized in table~\ref{tab:cons}.  In this case, the
prolongation $j^1 \txi$ of $\txi$ is the vector field along $\pi_{2,
  1}$ given in coordinates by
\[
  j^1\tilde{\xi} = \dot{x}_1 \pdd{x_1} + \dot{y}_1 \pdd{y_1}
  + \ddot{x}_1 \pdd{\dot{x}_1} + \ddot{y}_1 \pdd{\dot{y}_1}.
\]
The component $\JC_\oxi$ of the nonholonomic momentum map then
becomes
\[
  \JC_\oxi = j^1 \tilde{\xi} \contraction \Theta_L = m (\dot{x}_1^2 +
  \dot{y}_1^2),
\]
while the right-hand side of the momentum equation (\ref{momeq}) is
$\lie_{j^1 \tilde{\xi}} (L \d t) = m (\dot{x}_1 \ddot{x}_1 + \dot{y}_1
\ddot{y}_1) \d t$.  The momentum equation hence reduces to $\dot{x}_1
\ddot{x}_1 + \dot{y}_1 \ddot{y}_1 = 0$, which can also be verified
using the equations of motion.

\begin{table}
\begin{center}
\begin{tabular}{l|l}
  $(\alpha, \beta, \gamma, \delta)$ &  \\
  \hline \hline \vspace{-0.3cm}\\

  (1, 1, 0, 0) & $\dot{x}_1 \ddot{x}_1 + \dot{y}_1 \ddot{y}_1 = 0$ \\
  (0, 0, 1, 1) & $\dot{x}_2 \ddot{x}_2 + \dot{y}_2 \ddot{y}_2 = 0$ \\
  (1, 0, 0, -1) & $\dot{x}_1 \ddot{x}_1 - \dot{y}_2 \ddot{y}_2 = 0$
\end{tabular}
\caption{Conservation laws for the Benenti system.}\label{tab:cons}
\end{center}
\end{table}

\subsection{The nonholonomic Cosserat rod} \label{nonhcoss}

The nonholonomic Cosserat rod is an example of a nonholonomic field
theory studied in \cite{nonhcosserat}.  It describes the motion of a
rod which is constrained to roll without sliding on a horizontal
surface.  This theory can be studied using the bundle $\pi: Y
\rightarrow X$, where $X = [0, \ell] \times \bfR$ (space and time) and
$Y = X \times \bfR^2 \times \bfS^1$, with bundle coordinates $(s, t;
x, y, \theta)$.

Its Lagrangian is given by
\[
L = \frac{\rho}{2} (\dot{x}^2 +
  \dot{y}^2) + \frac{\alpha}{2} \dot{\theta}^2 - \frac{1}{2}\left(
    \beta (\theta')^2 + K \kappa^2 \right)
\]
with $\kappa^2 = (x'')^2 + (y'')^2$, while the constraints are given
by
\begin{equation} \label{constraints}
  \dot{x} + R \dot{\theta} y' = 0 \qqand \dot{y} - R \dot{\theta} x' =
  0.
\end{equation}
Here, $\rho$, $\alpha$, $\beta$, $K$, and $R$ are real parameters.
The field equations associated to this Lagrangian are given by
\begin{equation} \label{NHFE}
\left\{
  \begin{array}{rcl}
    \rho \ddot{x} + K x'''' & = & \lambda \\
    \rho \ddot{y} + K y'''' & = & \mu \\
    \alpha \ddot{\theta} - \beta \theta'' & = & R(\lambda y' - \mu x'),
   \end{array}
\right.
\end{equation}
where $\lambda$ and $\mu$ are Lagrange multipliers associated with the
nonholonomic constraints.  These equations are to be supplemented by
the constraint equations (\ref{constraints}).

For future reference, we note that the bundle of reaction forces is
generated in this case by the following forms:

\begin{equation} \label{constrforms}
\left\{ \begin{array}{rcl}
\Phi^1 & = & (\d x - \dot{x}\d t) \wedge \d s + Ry'(\d\theta -
\dot{\theta} \d t) \wedge \d s \\
\Phi^2 & = & (\d y - \dot{y}\d t)
\wedge \d s - Rx'(\d\theta - \dot{\theta} \d t) \wedge \d s.
\end{array} \right.
\end{equation}

In the absence of nonholonomic constraints, this model is subject to
the usual symmetry actions such as translations in time, global
translations, and global rotations (see \cite{nonhcosserat}).  As we
shall now show, some of these persist in the nonholonomic case.

Observe that the Lagrangian is of second order.  At the cost of
sacrificing physical relevance, one may also put $K = 0$ to obtain a
purely first-order theory.  We will not do so here, as the derivation
of a nonholonomic momentum lemma for a second-order field theory
proceeds exactly as above, up to a few minor modifications.  The
nonholonomic momentum map $\JC$ is now defined on $J^3 \pi$, and the
nonholonomic momentum equation hence becomes
\begin{equation} \label{eqmomnon}
  (j^3\phi)^\ast (\d \JC_{\bar{\xi}}) =
  (j^2\phi)^\ast \lie_{j^2 \tilde{\xi}}(L \eta),
\end{equation}
where $\phi$ is a solution of the nonholonomic Euler-Lagrange
equations.  This equation holds both for ``genuine'' symmetries as
well as for symmetries along the projection.

\subsubsection{Translations in time}

Consider the action of $\bfR$ on $X$ by translations in time defined
by the map $\Phi: \bfR \times X \rightarrow X$, with $\Phi(\alpha, (s,
t)) = (s, t+\alpha)$.  As the bundle $\pi$ is trivial, this action
naturally induces an action on $Y$, and, by prolongation, also on
$J^1\pi$.  Clearly, this action is not vertical; the fundamental
vector field associated to a Lie algebra element $\xi \in \bfR$ is
given by
\begin{equation} \label{vf}
  j^1 \xi_Y = \xi \pdd{t}.
\end{equation}

This vector field is a nonholonomic symmetry: it is $\pi_1$-related to
a vector field on $X$ and its contraction with the elements of $F$
vanishes along $\calC$: for the $n$-form $\Phi^1$ defined in
(\ref{constrforms}), we have
\[
\frac{\partial}{\partial t} \contraction \Phi^1 = - (\dot{x} +
R\dot{\theta}y') \d s,
\]
which vanishes on $\calC$, and a similar result holds for $\Phi^2$.

Note that the nonholonomic symmetry (\ref{vf}) is a horizontal
symmetry (definition~\ref{def:hsym}); in general, this will not be the
case.  In this special case, however, we have that $\lie_{j^1 \xi_Y}
(L \eta) = 0$, expressing the infinitesimal invariance of the
Lagrangian.  The momentum map now becomes
\[
\fl  (j^3\phi)^\ast \JC_1 =
  \left[  - K x'''\dot{x} + - Ky'''\dot{y}
    + \beta \theta' \dot{\theta} + K(x'' \dot{x}' + y'' \dot{y}')
  \right] \d t
   + \calE \d s \nonumber,
\]
where we have introduced the \emph{energy density}
\[  \calE =
      \frac{\rho}{2}(\dot{x}^2 + \dot{y}^2) +
      \frac{\alpha}{2} \dot{\theta}^2 + \frac{K}{2}((x'')^2 + (y'')^2) +
      \frac{\beta}{2} (\theta')^2.
\]
The nonholonomic momentum lemma then states that $\d (j^3 \phi)^\ast
\JC_1 = 0$. This equation expresses local conservation of energy; by
integrating over a hypersurface of constant $s$, we may then obtain a
law expressing global conservation of energy.

\begin{remark}
  This case was also treated in \cite{nonhcosserat} by use of a
  different method, to which we refer for further details.  Note
  however, that the method used in that paper is less general as it is
  only valid for horizontal nonholonomic symmetries.
\end{remark}

\subsubsection{Spatial translations} \label{sec:spattr}

Consider the action of $\bfR^2 \times \bfS^1$ on $Y$ by translations;
\emph{i.e.} for each $(a, b, \varphi)$ we consider the map $\Phi_{(a, b,
  \varphi)}: (s, t, x, y, \theta) \mapsto (s, t, x + a, y + b, \theta
+ \varphi)$.  Let $\xi = (v_1, v_2, v_\theta)$ be an element of the
Lie algebra of $\bfR^2 \times \bfS^1$.  The corresponding fundamental
vector field is given by
\[
  \xi_Y = v_1 \pdd{x} + v_2 \pdd{y} + v_\theta \pdd{\theta}.
\]

When no constraints are present, this symmetry implies the
conservation of linear momentum.  In the presence of nonholonomic
constraints, a modified conservation law holds: it is easy to see that
the following vector field annihilates $F$ along $\calC$:
\[
  \tilde{\xi} = - Ry' \pdd{x} + Rx' \pdd{y} + \pdd{\theta}.
\]
(Any scalar multiple of the above vector field is also allowed.)
This generalized vector field corresponds with the section
$\bar{\xi} = (-Ry', Rx', 1)$ of $\pi_{1, 0}^\ast \gothg^F$. As
$\txi$ is vertical, the nonholonomic momentum lemma
\ref{thm:nhmomgen} can be applied.

The right-hand side of (\ref{eqmomnon}) is
\[
  j^2\phi^\ast \lie_{\tilde{\xi}}(L
  \eta) = \big[ -R\rho\dot{y}'\dot{x} + R\rho\dot{x}'\dot{y}
    - KR x''' y'' + KRy'''x'' \big] \eta.
\]
The nonholonomic momentum map $\JC$, on the other hand, is given by
\[
  \JC_{\bar{\xi}}  = -\big[\rho(Rx'\dot{y} - Ry'\dot{x}) +
    \alpha\dot{\theta} \big] \d s
  - \big[ KR(y'x''' - x'y''') + \beta \theta' \big] \d t,
\]
and the nonholonomic momentum equation hence becomes
\begin{equation} \label{nomults}
  Ry'(\rho \ddot{x} + K x'''') - Rx'(\rho \ddot{y} + K y'''') =
    \alpha \ddot{\theta} - \beta \theta''.
\end{equation}
This conservation law can also be derived from the nonholonomic field
equations (\ref{NHFE}) by subtracting the second equation multiplied
by $x'$ from the first equation multiplied by $y'$, and using the
third equation to eliminate the Lagrange multipliers $\lambda$ and
$\mu$.  Unfortunately, the knowledge of this nonholonomic conservation
law does not help us in solving the field equations (in contrast with
the situation for the vertical rolling disc; see \cite{bloch}).


\section{Conclusions}

In this paper, we presented a geometric framework for nonholonomic
field theories with symmetries.  We showed that the momentum map
associated with a group action satisfies a certain momentum equation,
which we proved in a number of cases.  On the one hand, there exists a
momentum equation for nonholonomic symmetries which act nontrivially
on the base space (examples being energy conservation for the
nonholonomic rod), while on the other hand a similar result exists for
generalized symmetries associated to a vertical group action.  The
Benenti system is an example of the latter.

It is likely that the results in this paper can be generalized still
further.  In particular, there seems to be no reason why there
shouldn't be a momentum equation for generalized symmetries associated
to a non-vertical group action, up to some technical restrictions (for
instance, that the generalized vector field should be related to a
regular vector field on the base space).  This generalization would
encompass both momentum equations derived in this paper.  Moreover, it
would be interesting to study further examples of such a momentum
equation.

\ack

The authors would like to thank F. Cantrijn and D. Saunders for
stimulating discussions and comments.

The first author is a Postdoctoral Fellow from the Research Foundation
-- Flanders (FWO-Vlaanderen), and a Fulbright Research Scholar at the
California Institute of Technology.  Additional financial support from
the Fonds Professor Wuytack is gratefully acknowledged.

The second author is supported by MEC (Spain) Grants MTM 2004-7832 and
MTM2007-62478, project ``Ingenio Mathematica'' (i-MATH) No. CSD
2006-00032 (Consolider-Ingenio 2010) and S-0505/ESP/0158 of the CAM.

\section*{Appendix: noncovariant nonholonomic constraints}
\setcounter{section}{1}
\setcounter{subsection}{0}
\setcounter{definition}{0}


A special class of nonholonomic field theories consists of those where
the base space $X$ is $\bfR \times M$, where the first factor
represents time, and such that $\pi$ is trivial.  In other words,
there exists a canonical distinction between time and space.
Accordingly, one can show that in this case, the jet bundle $J^1\pi$
is isomorphic to the product bundle $\bfR \times [J^1(M, S) \times_S
TS]$, thus providing a canonical distinction between derivatives of
the fields with respect to time and space.

This class of field theories was discussed in detail in
\cite{nonhcosserat} and includes among others also theories of
nonrelativistic elasticity (see also \cite{MPSW01}).  The example
studied in section~\ref{nonhcoss} is also among these field theories.

As shown in \cite{nonhcosserat}, the reaction forces for this kind of
field theory are not the ones obtained from the Chetaev prescription
in remark~\ref{rem:chetaev}.  The problem is that this form of the
principle is fully covariant, in the sense that no distinction is made
between spatial derivatives and derivatives with respect to time.
However, in some cases (the nonholonomic Cosserat rod being one of
them) the time derivatives \emph{do} play a distinguished role, and
therefore a different, ``noncovariant'' Chetaev principle is needed.

The main objective of this appendix is to propose such a principle,
using the geometry of $J^1\pi$, and especially the isomorphism with
$\bfR \times [J^1(M, S) \times_S TS]$.  Using this modified principle,
we derive the correct form of the reaction forces for the nonholonomic
Cosserat rod, which were obtained in \cite{nonhcosserat} using a
different approach.

\subsection{A new vertical endomorphism} \label{sec:vertnc}

Recall the coordinate expression (\ref{vertend}) of the vertical
endomorphism $S_\eta$ on $J^1\pi$.  This tensor field was constructed
by Saunders \cite{Saunders89} using a map assigning to each one-form
$\omega$ on $X$ the vector-valued one-form $S_\omega$ on $J^1\pi$
given in coordinates by (see \cite[p.~156]{Saunders89})
\begin{equation} \label{Seenvorm}
  S_\omega = \omega_\mu (\d{y^a} - y^a_\nu \d{x^\nu}) \otimes
  \pdd{y^a_\mu}, \where \omega = \omega_\mu \d x^\mu.
\end{equation}

Roughly speaking, the vertical endomorphism $S_\eta$ then arises,
once a volume form on $X$ is chosen, by putting
\begin{equation} \label{wedgeendo}
  S_\eta = S_{\d{x^\mu}} \dwedge (\pi_1^\ast \d^nx_\mu),
\end{equation}
where the wedge operator `$\dwedge$' is defined as follows: if $\Phi$
is a vector-valued $k$-form on $J^1\pi$, and $\alpha$ is a regular
(\emph{i.e.} $\bfR$-valued) $l$-form, then $\Phi \dwedge \alpha$ is the
vector-valued $(k + l)$-form given by $\dual{\Phi \dwedge \alpha,
  \beta} = \dual{\Phi, \beta} \wedge \alpha$ for all $\beta \in
\Omega^1(J^1\pi)$.

It is obvious that (\ref{wedgeendo}) is fully covariant, in the sense
that no distinction is made between the variables on the base space.
However, in elastodynamics, this is not always desirable, as we have a
distinguished direction of time.  Therefore, we propose the following
``non-covariant'' vertical endomorphism:
\begin{definition} \index{vertical endomorphism!noncovariant}
  The \emph{non-covariant vertical endomorphism} is the vector-valued
  $(n + 1)$-form $\Snc$ defined as $\Snc := S_{\d{t}} \dwedge
  (\pi_1^\ast \eta_M)$, where $S_{\d{t}}$ is the vector-valued
  one-form associated to $\d{t}$ as in (\ref{Seenvorm}).
\end{definition}
Here $\eta_M$ denotes the volume form on $M$ defined by
restriction to $M$ of $\d^{n}x_0$

Note that $\d{t}$ is a well defined one-form on $X = \bfR \times M$;
therefore, $\Snc$ is an intrinsic object.  In coordinates, $\Snc$ is
given by
\[
  \Snc = (\d y^a - y^a_\mu \d x^\mu) \wedge \d^{n}x_0
    \otimes \frac{\partial}{\partial y^a_0}.
\]

\begin{remark}
  In \cite{Saunders89}, it is shown that the action of the vertical
  endomorphism $S_\eta$ is related to the fact that $\pi_{1, 0}: J^1
  \pi \rightarrow Y$ is an affine bundle; \emph{i.e.}  there exists an
  affine action of $\pi^\ast T^\ast X \otimes V \pi$ on $J^1\pi$.  The
  noncovariant vertical endomorphism can be understood in a similar
  vein, by restricting this affine action to $\pi^\ast T^\ast \bfR
  \otimes V \pi$.
\end{remark}

\subsection{The bundle of constraint forces} \label{sec:CF}

Let $\iota: \calC \hookrightarrow J^1\pi$ be a constraint manifold.
In section~\ref{sec:1stFE}, we required that reaction forces be
$n$-horizontal and $1$-contact.  This leads to local expressions of
the form (\ref{locform}).  For noncovariant constraints, we attribute
a special status to the time coordinate, and therefore we require also
that the following holds:
\[
  i_v i_w \Phi = 0
\]
for all tangent vectors $v, w$ on $J^1 \pi$ such that $T (\pr_1 \circ
\pi_1)(v) = T(\pr_1 \circ \pi_1)(w) = 0$, where $\pr_1 : \bfR \times M
\rightarrow \bfR$ is the projection onto the first factor (this
condition expresses that $v$ and $w$ do not contain a component
proportional to $\frac{\partial}{\partial t}$).  In coordinates, this
implies that $\Phi$ has the following form:
\begin{equation} \label{noncovform}
  \Phi = A_a (\d{y^a} - y^a_\mu \d{x}^\mu) \wedge \d^n x_0,
\end{equation}
where the $A_a$ are local functions on $\calC$.  Compare this with
(\ref{locform}) and note that only the ``time'' component is
left (\emph{i.e.} the component proportional to $\theta^a \wedge (
\frac{\partial}{\partial t} \contraction \eta)$).

Using the noncovariant vertical endomorphism of
section~\ref{sec:vertnc}, one can construct from $\calC$ a natural
candidate for the bundle $F$.  This is a field-theoretic
generalization of the well-known Chetaev principle from mechanics.

Recall that $\calC$ is assumed to be given by the vanishing of $k$
functions $\varphi^\alpha$ and define the associated bundle of
reaction forces as the bundle $F$ locally spanned by the following $(n
+ 1)$-forms: $\Phi^\alpha := \Snc^\ast( \d\varphi^\alpha)$, or in
coordinates:
\begin{equation} \label{noncovphi}
  \Phi^\alpha = \frac{\partial \varphi^\alpha}{\partial y^a_0}
    (\d{y}^a - y^a_\mu \d{x}^\mu) \wedge \d^nx_0.
\end{equation}
The $(n + 1)$-form $\Phi^\alpha$ is therefore of the form outlined in
(\ref{noncovform}), with $A^\alpha_a = \frac{\partial
  \varphi^\alpha}{\partial y^a_0}$.

\section*{References}


\providecommand{\arxiv}[1]{\texttt{arXiv:#1}}\providecommand{\href}[1]{\texttt%
{#1}}

\end{document}